\documentclass[twocolumn,elsarticle,preprintnumbers,amsmath,amssymb]{revtex4}

\usepackage{xspace}
\usepackage{graphicx}
\usepackage{dcolumn}
\usepackage{bm}

\def\g{\gamma}
\def\beq{\begin{equation}}
\def\eeq{\end{equation}}
\def\bea{\begin{eqnarray}}
\def\eea{\end{eqnarray}}
\def\bmat{\begin{pmatrix}}
\def\emat{\end{pmatrix}}
\def\bei{\begin{itemize}}
\def\eei{\end{itemize}}

\begin{document}

\preprint{MCTP-11-36}
\preprint{CERN-PH-TH/2011-041}

\title{Higgs boson search significance deformations due to mixed-in scalars}

\author{Rick S. Gupta}\email{guptasr@umich.edu}
\author{James D. Wells}\email{jwells@umich.edu}
\affiliation{CERN Theoretical Physics (PH-TH), CH-1211 Geneva 23, Switzerland}
\affiliation{MCTP, Department of Physics, University of Michigan, Ann Arbor, MI 48109}

\date{\today}

\begin{abstract}
The existence of exotic scalars that mix with the Standard Model (SM) Higgs boson can affect Higgs boson phenomenology in a multitude of ways. We consider two light Higgs bosons with shared couplings to SM fields and with masses close to each other,  in the range where the $h \to WW \to l \nu l \nu$ is an important search channel. In this channel, we do not find  the dilution of significance of the `SM-like'   Higgs boson that is naively expected because of the mixing.  This is because of leakage of events from the other scalar into its signal region. Nevertheless, we show that the broadening of the $h\to WW \to l \nu l \nu$ significance plots of Standard Model Higgs boson searches could indicate the first evidence of the  the extra scalar state. 
\end{abstract}
\maketitle

{\it New mixed-in scalars.}
We want to consider cases where the  Standard Model (SM) Higgs mixes with  scalar  states  leading to multiple Higgs bosons with ``shared" couplings to SM fields. The most natural choice for the  extra scalar is a singlet but doublets and triplets that do not get a vacuum expectation value (VEV) are also possible candidates.   There are many reasons to consider extra scalar particles that are singlets under the SM gauge groups. The existence of these exotic new particles, perhaps from a hidden sector, are of particular phenomenological significance. This is because a relevant, gauge-invariant and Lorentz invariant operator, $|\Phi_i|^2$, can be formed by them, thereby   enabling a simple renormalizable coupling with the SM Higgs boson: $|H_{SM}|^2|\Phi_i|^2$, where $H_{SM}$ is the SM Higgs boson and $\Phi_i$ are exotic scalar states. If $\Phi_i$ gets a VEV the SM Higgs boson mixes with $\Phi_i$ and Higgs phenomenology is no longer SM Higgs phenomenology, but one of multiple scalar states sharing couplings to SM fields according to the strength of the wave-function overlap with the SM Higgs boson.  A more complete discussion of the theory can be found in~\cite{Wells}. While complex scalars charged under new hidden gauge groups are of more interest to us, our analysis holds also for a real scalar, $\phi$. In this case the interaction term $|H_{SM}|^2\phi$ will also contribute to the mixing between the scalars.

Such sharing of couplings can also arise from the mixing of the SM neutral Higgs component with the neutral components of an exotic doublet, $H'$, or a triplet with no hypercharge, $\Sigma$, provided $H'$ or $\Sigma$ get a vanishing or  small VEV.   A triplet is, of course,  required to have no VEV as it would otherwise give tree-level contributions to the $\rho$-parameter.  As discussed in~Ref.~\cite{Gupta:2009wn}, vanishing or small VEVs are favored by  experimental constraints  for exotic doublets too~\footnote{As the exotic VEV,  $v_2\to 0$, and hence $\beta=\tan^{-1} (v_2/v_1) \to 0$,  the mixing angle, $\alpha$, between the mass eigenstates $h$ and $H$, which is proportional to $\beta$ in this limit, also vanishes, i.e. $\alpha= c \beta \to 0$. However, depending on the parameters of the potential, we can still, have $\alpha/\beta$ relatively large (for the case of the doublets, see for eg., eq. 88 in  Ref.~\cite{Gupta:2009wn}) so that we  can have a non-negligible mixing angle  $\alpha$ even if $\beta$ is much smaller.}.


In this letter we wish to show how the standard significance searches for the SM Higgs boson are affected by the existence of these mixed-in states. Careful inspection of the $h\to WW \to l \nu l \nu$ ``significance plots" could reveal the existence of two more more Higgs bosons, even before the resonance of the second Higgs boson has been found in another search channel with better mass resolution.

We will first describe the method of making significance projections for the SM Higgs boson. For illustration we will proceed along the lines of the ATLAS analysis which has a similar sensitivity to the CMS study but is much easier to reproduce. We will comment on the CMS study at the end.  We will then describe how the existence of an extra mixed-in scalar state would alter the significance plots, showing that new physics could be revealed through that shape first.  And finally we will make some concluding comments.

\medskip
{\it LHC Sensitivity Projections.}
We will first reproduce the ATLAS sensitivity projections for Higgs searches at the LHC for 7 TeV center of mass energy that were  made in Ref.~\cite{atlas2}. We  concentrate on the $h\to WW \to l \nu l \nu$  ($l= e,\mu$) channel as this is the most sensitive channel in the range $125-190$ GeV.  In the range $130- 180$ GeV this is by far the  dominant channel and sensitivity limits obtained from just this channel alone are very close to limits obtained by combining all channels. For $m_h\lesssim130$ GeV the $h \to \g\g$ channel starts to become competitive with the $h\to WW \to l \nu l \nu$ channel  and for $m_h \gtrsim 190$ GeV the  $h \to ZZ \to 4l$ channel becomes important so that considering the $h\to WW \to l \nu l \nu$ channel alone  for these masses would give us weaker  sensitivity estimates compared to estimates evaluated by combining all channels. 

For our computations  we will use the expected Standard Model (SM) signal ($S_{SM}$) and background ($B$) values for  ${\cal L}=1~$fb$^{-1}$ integrated luminosity given in Ref.~\cite{atlas2}. The values for $S_{SM}$ and $B$ for the $h\to WW \to l \nu l \nu$ channel have been given in Ref.~\cite{atlas2} as a function of  the putative mass of the Higgs $m_h$ used for the  search. The only $m_h$-dependent cut   that has been applied in Ref.~\cite{atlas2}  is,
\beq
m_T \leq m_h
\label{cut}
\eeq
where the transverse mass $m_T$ is defined by $m_T= \sqrt{(E_T^{ll}+E_T^{miss})^2-(\textbf{P}_T^{ll}+E_T^{miss})^2}$,  the transverse momentum of the lepton pair,  $\textbf{P}_T^{ll}=\textbf{P}_T^{l1}+\textbf{P}_T^{l2}$, $E_T^{ll}=\sqrt{P_T^{ll~2}+m_{ll}^2}$,  $E_T^{miss}$ is the missing transverse energy and $m_{ll}$ is the invariant mass of the leptons~\cite{Barr:2009mx}. This cut utilizes the fact that for a Higgs decaying to $l \nu l \nu$, $m_T$ is always less than the Higgs mass. The other cuts used in Ref.~\cite{atlas2} have been described in more detail in Ref.~\cite{atlas0} .

\begin{figure}[t]
\begin{center}
\includegraphics[width=0.7\columnwidth]{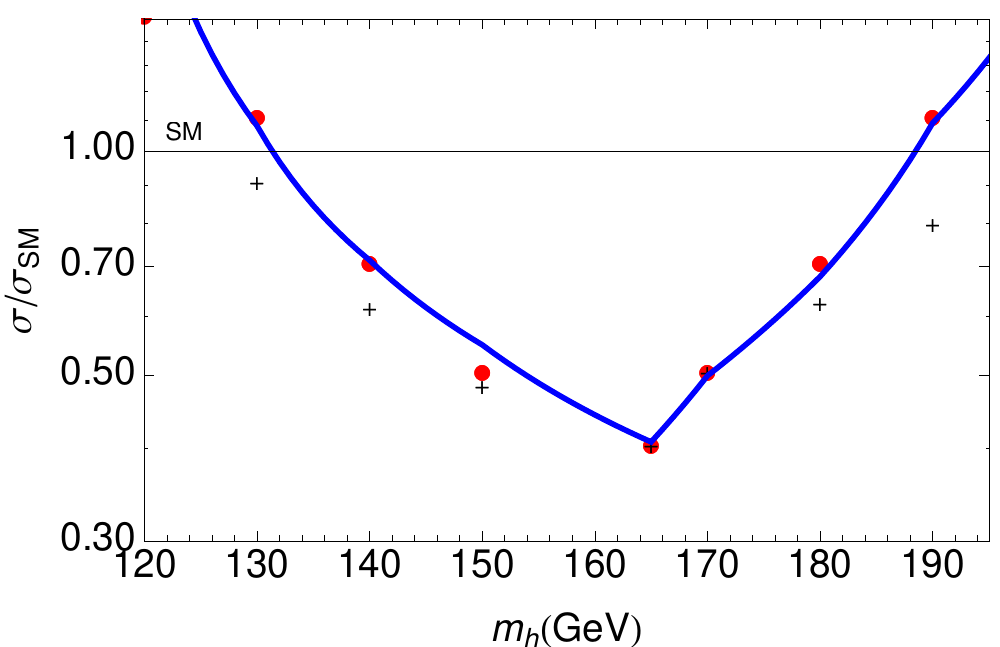}
\end{center}
\caption{The multiple of the cross-section of a Standard Model Higgs boson which can be excluded at 95$\%$ confidence level using  1 fb$^{-1}$ LHC data at 7 TeV by the $h \to WW \to l\nu l\nu$ channel. For comparison we also show the ATLAS projections that appear in Ref.~\cite{atlas2} for the $h \to WW \to l\nu l\nu$ channel (red dots)  and after combining all channels (``+" signs).}
\label{proj}
\end{figure}

For evaluating exclusion confidence levels and discovery significances we use simple event counting estimates  assuming a Gaussian distribution for the expected number of  events. We   review the  procedure for setting exclusion limits and finding significances in detail in the appendix. For the exclusion estimates we use the fact that  a signal value $S$ still allowed after applying  95$\%$ confidence level exclusion bounds must satisfy,
\beq
\frac{S}{\sqrt{S+B+ (\Delta B)^2}}  \leq 1.64. 
\label{exc}
\eeq
Here $\Delta B$ is the systematic error. As far as significance estimates are considered we use the significance estimator (defined as $S_{c12}$ in~\cite{Ball:2007zza}),
\beq
Z_0 = 2(\sqrt{S+B}-\sqrt{B})\sqrt{\frac{B}{B+ (\Delta B)^2}}.
\eeq
Taking $\Delta B/B =0.15$  we find projections for the $95\%$ upper limit on $S/S_{SM}$ that can be put with ${\cal L}=1$ fb$^{-1}$.   As values of the signal and background  have been given only for a discrete set of masses in Ref.~\cite{atlas2}, for intermediate masses we have linearly interpolated. The results, shown in Fig.\ref{proj}, agree very well with ATLAS projections for the reach of the $h\to WW \to l\nu l\nu$ channel (red dots in Fig.\ref{proj}). We also show in Fig.\ref{proj} the projected ATLAS limits  obtained in Ref.~\cite{atlas2} after combining all the channels.  The numerical value $\Delta B/B =0.15$ has been chosen to get maximum agreement with  the ATLAS projections in Ref.~\cite{atlas2}. As mentioned earlier after combining all the channels stronger limits can be obtained although the limits from the $h \to WW\to l \nu l \nu$ channel alone are close to the combined limits for $130{\rm~GeV}<m_h<190{\rm~GeV}$.

\medskip
{\it Significance with mixed-in Higgs bosons.}
As can be seen from Fig.\ref{proj} even if the SM Higgs is excluded at a certain mass it is still possible to have a Higgs boson at that mass if the production cross-section times branching ratio is  suppressed by $\xi$ compared to the SM. We want to consider the scenario where there are two Higgs scalars, $h$ and $H$, and the production cross-section times branching ratio of these Higgs bosons is suppressed as follows,
\bea
\sigma_h=\xi \sigma_{SM}(m_h) \nonumber\\
\sigma_H=(1-\xi) \sigma_{SM}(m_H),
\label{cs}
\eea
where $\xi\leq 1$. For a sufficiently large $\xi$ it would be hard to detect the scalar $H$.  In a situation where only $h$ is detected, deviations from SM can still be detected if the cross-section for production of $h$, which would be smaller than  the SM expectation, can be measured. As we will see, however, it may not always be easy to detect such a deviation in the cross-section.

\begin{figure}
\begin{center}
\includegraphics[width=0.75\columnwidth]{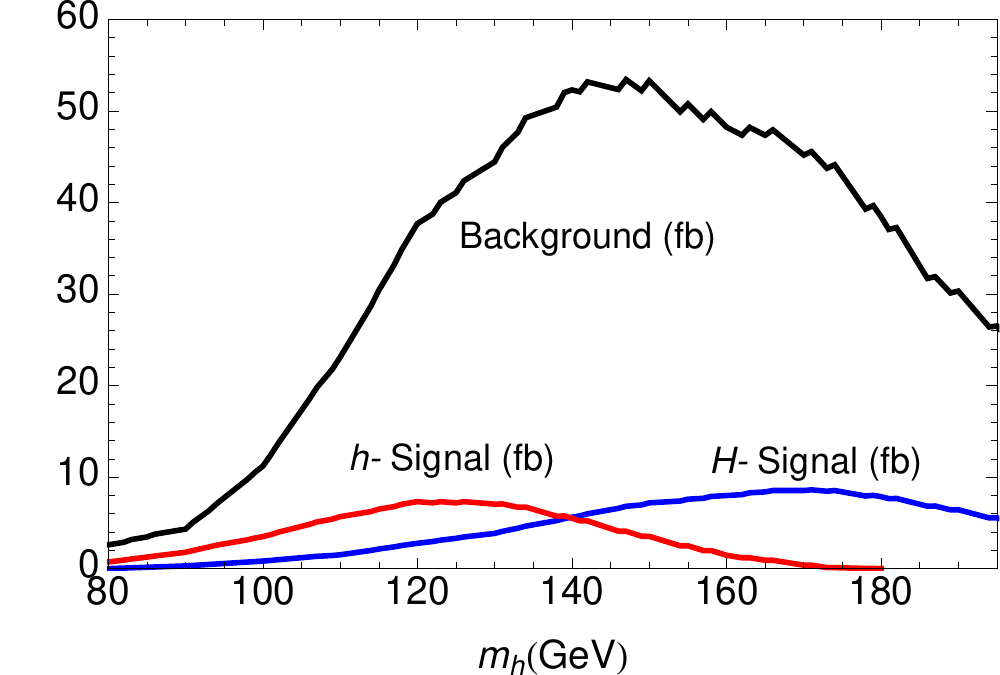}

\end{center}
\caption{The signal cross-section for $h$ and $H$ with $m_h^{true}=125$ GeV, $m_H^{true}=170$ GeV and $\xi=0.8$ and the  SM background cross-section after applying the cut in eq.~\ref{slide}. The signal here is that of a Standard Model Higgs with mass 170 GeV. }
\label{bgr}
\end{figure}

We will now describe how the $h\to WW \to l \nu l \nu$ significance plots are distorted for the scenario mentioned above. An important difference from the  last section is that instead of using the cut in eq.~\ref{cut}  for the search we use the sliding mass window,
\beq
0.75~m_h<m_T<m_h.
\label{slide}
\eeq
This is the  cut being used by ATLAS in their present searches~\cite{armbuster,Aad:2011qi,ATLAS:2011aa}. The background $B$ after applying the above cut can be easily calculated from the background values given in Ref.~\cite{atlas2} where the cut in eq.~\ref{cut} has been applied, by using \footnote{In Ref.~\cite{atlas2}  background values for $B (m_T< m_h)$ for $m_h<120$ GeV have not been provided; for $m_h<120$ GeV we use the shape of the $m_T$-distribution curves for the background provided in~\cite{armbuster} keeping the normalization of Ref.~\cite{atlas2}. Note that~\cite{armbuster} considers the signal and background only for the $H+0$ jet and $H+1$ jet  analyses whereas Ref.~\cite{atlas2} also consider the subdominant  $H+2$ jet contribution.}
\bea
& B(0.75~m_h<m_T<m_h) =  & \nonumber \\
 & B (m_T< m_h)- B (m_T< 0.75~m_h). &
\eea
We show in Fig.\ref{bgr} the background obtained by applying the  cut in eq.~\ref{slide}.  One can similarly reconstruct the background  $m_T$-distribution. For a particular $m_h$-bin, we will get a cross-section equal to,
\bea
B (m_T< m_h)- B (m_T< m_h-10{\rm GeV)}.
\label{subtle}
\eea

\begin{figure}[t]

\begin{center}
\includegraphics[width=0.7\columnwidth]{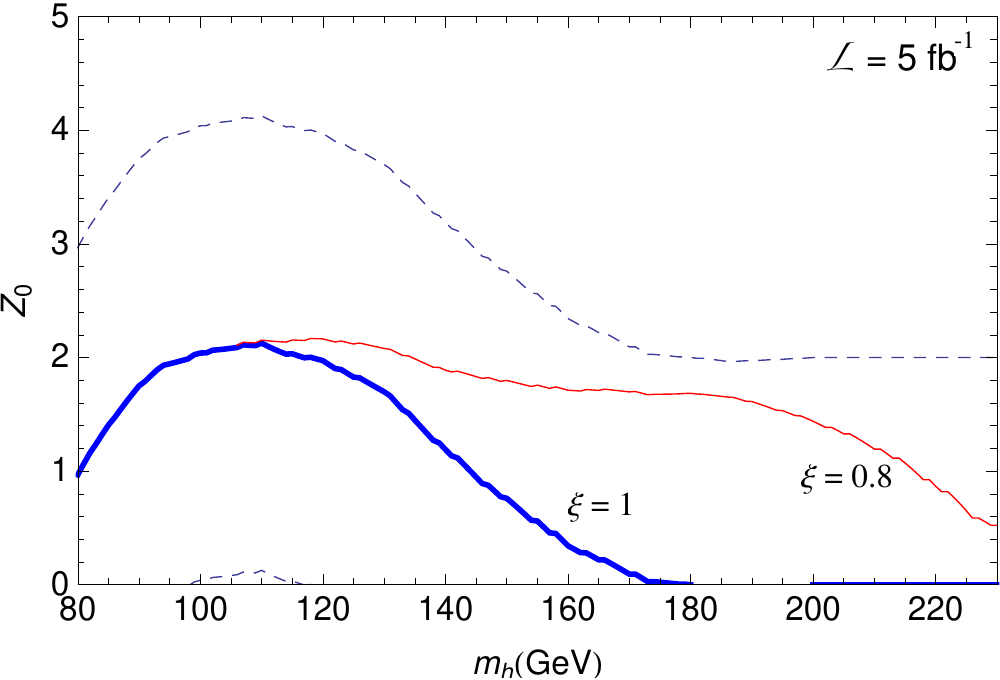}
\end{center}
\caption{The significance vs $m_h$ curve for 5 fb$^{-1}$ data after applying the sliding mass cut in eq.~\ref{slide} on  the signal and background  assuming $m_h^{true}=125$ GeV and $m_H^{true}=170$ GeV. We show the curve for $\xi=0.8$ and $\xi=1$. The dashed line shows the 2-sigma band around the $\xi=1$ line.}
\label{5fb}
\end{figure}

There is a subtlety which must be kept in mind if this cut is used for discovery searches and not just for setting exclusion limits. If a Higgs boson does exist at a certain mass $m_{h/H}^{true}$, we will obtain a significance after applying the sliding mass cut above, even for Higgs masses different from $m_{h/H}^{true}$. Thus instead of a sharp peak in significance  at  $m_{h/H}^{true}$  a broad excess would be seen around $m_{h/H}^{true}$, and the peak significance would not necessarily be obtained at  $m_{h/H}^{true}$ if the background after applying the above cut is not flat  with respect to $m_h$.  As shown in Fig.\ref{bgr} the background rises for $m_h\lesssim 135$ GeV and falls for $m_h\gtrsim 150$ GeV. Consider the case of  a Higgs boson with  $m_h^{true}=125$ GeV. Although the signal is maximum if one takes $m_h= 125$  GeV in the sliding window in eq.~\ref{slide}, the background is  smaller for lower values of $m_h$, as shown in Fig.\ref{bgr}, so that the maximum significance is obtained at a mass lower than the 125 GeV. This can be seen from $\xi=1$ curve in Fig.\ref{5fb} that shows the significance vs.\ $m_h$ curve peaking below 125 GeV. Note, the plot was made for $\xi=1$ but it would have the same shape (i.e., same peak position) for any $\xi$ value. On the other hand,  the significance curve for a 170 GeV Higgs would peak at values higher than 170 GeV because the background falls for $m_h > 170$ GeV.   



Before discussing an example we mention how the significances scale with integrated luminosity. Because of systematic effects  the significances do not scale as ${\cal L}^{0.5}$ but as  ${\cal L}^\alpha$ where $\alpha$ varies between 0.3 and 0.6~\cite{atlas2}.  In this work we  take $\alpha=0.4$ throughout.


\begin{figure}[t]
\begin{center}
\includegraphics[width=0.76\columnwidth]{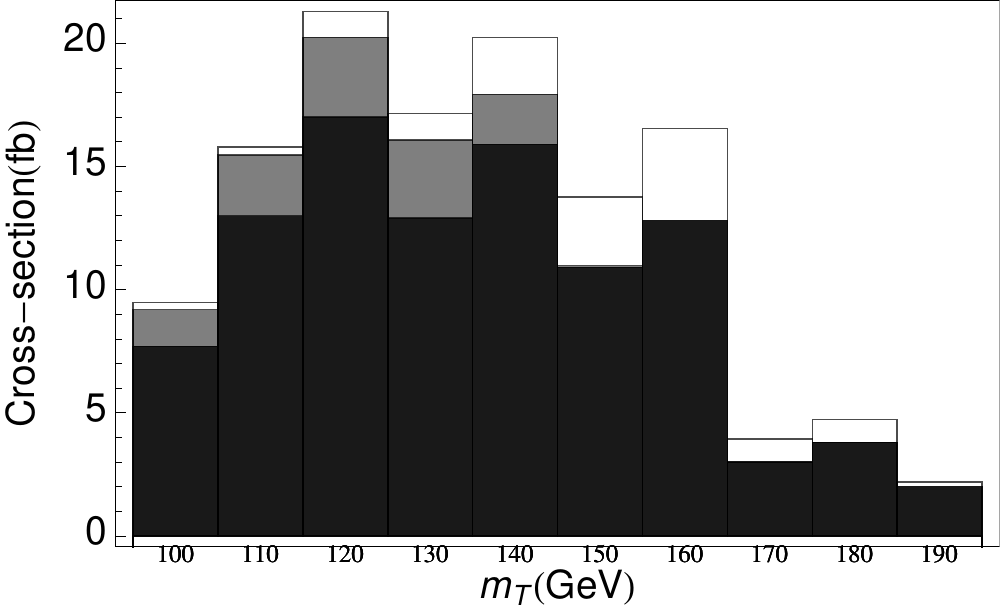}
\end{center}
\caption{ We show the $m_T$-distributions for the background (black)  and the signal plus background  with  $\xi =1$ (grey) and $\xi= 0.8$ (white). We have taken $m_h^{true}=125$ GeV and $m_H^{true}=170$ GeV.  In the $\xi=0.8$ case the effect of the 170 GeV Higgs can be  seen in the presence of the excess for $m_h>125$ GeV.  }
\label{bars}
\end{figure}

\begin{figure}[t]
\begin{center}
\includegraphics[width=0.7\columnwidth]{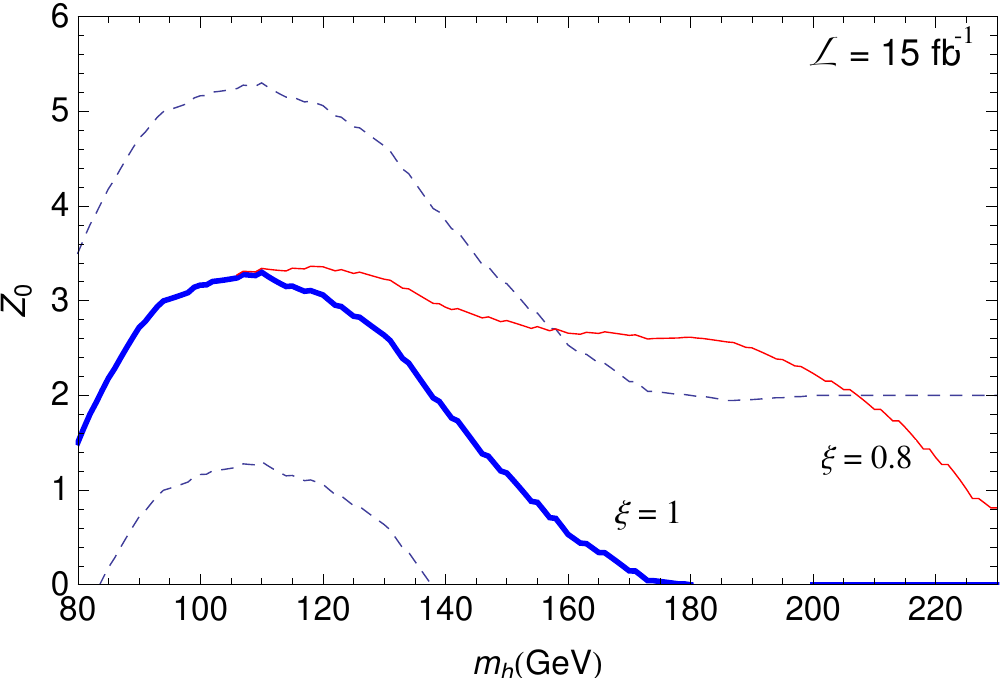}
\end{center}
\caption{The significance vs $m_h$ curve for 15 fb$^{-1}$ data after applying the sliding mass cut in eq.~\ref{slide} on  the signal and background  assuming $m_h^{true}=125$ GeV and $m_H^{true}=170$ GeV. We show the curve for $\xi=0.8$ and $\xi=1$. The dashed line shows the 2-sigma band around the $\xi=1$ line.}
\label{10fb}
\end{figure}
\begin{figure}[t]
\begin{center}
\includegraphics[width=0.7\columnwidth]{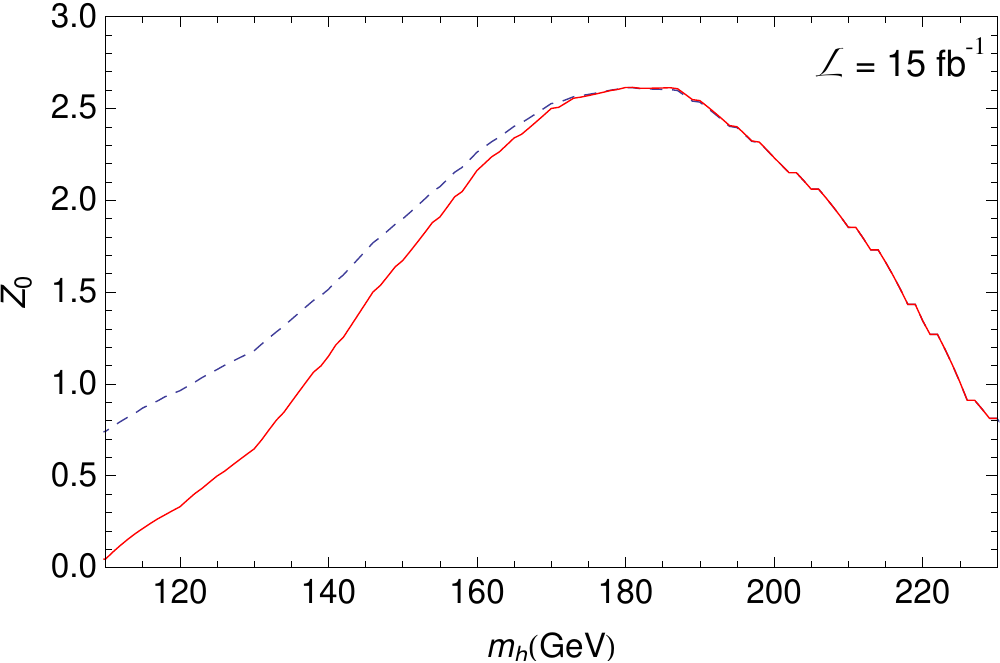}
\end{center}
\caption{The significance vs $m_h$ curve for 15 fb$^{-1}$ data after applying the sliding mass cut in eq.~\ref{slide} on  the signal and background,  assuming $m_h^{true}=125$ GeV, $m_H^{true}=170$ GeV, $\xi = 0.8$  and treating  the contribution of a SM Higgs at 125 GeV as background.  The dashed curve shows the  significance curve if   the correct light Higgs contribution with  $\xi=0.8$ is subtracted from the signal and included in the background.}
\label{sub}
\end{figure}
{\it Example with one extra mixed-in Higgs boson.}
We want to illustrate the distortions in the significance plots  that arise if there are two Higgs bosons with cross-sections given by eq.~\ref{cs} but  the standard single Higgs search strategy is used. In order to better understand the significance profiles we need to look at the underlying $m_T$-distributions for the signal and background first. We take $m_h^{true}=125$ GeV and $m_H^{true}=170$ GeV and consider  two cases with  $\xi =1$ and $\xi= 0.8$. For the $m_T$-distribution of the signal we take the shape from Ref.~\cite{Barr:2011ux} and Ref.~\cite{Aad:2011qi}  for $h$ and $H$ respectively and we use the signal cross-section values provided in Ref.~\cite{atlas2} for the normalization. We show the $m_T$-distributions for the signal plus background taking   $\xi =1$ and $\xi= 0.8$  in Fig.\ref{bars}. As one can see even though the production cross-section for $h$ is smaller in the $\xi=0.8$ case, the $m_T$ distribution  hardly changes from the $\xi=1$ case (the SM limit)  in the  $m_h<125$ GeV region of the plot. The reason is that for the $\xi=0.8$ case, there are extra events from the decay of the scalar $H$  having $m_T<$ 125 GeV. As the SM cross-section  in this channel at 170 GeV is much higher  than the cross-section  at 125 GeV this leakage of  $H$-decay events into the  $m_h<125$ GeV region is substantial.  For $m_h>125$ GeV there is a noticeable difference in the $\xi=0.8$ case as there is now an excess  in this region which is not expected for $\xi=1$. Such an  excess can, however, easily be misinterpreted as an error in background modeling or a background fluctuation. If such an excess persists, to ascertain whether it is due to a mixed-in scalar or  a background modeling error, would require more effort in reanalyzing and understanding the different backgrounds. To quantify the significance of this excess we need to look at the significance plots  shown in Fig.\ref{5fb} and Fig.\ref{10fb}.  As can be seen from Fig.\ref{5fb} and Fig.\ref{10fb} there is no noticeable feature at 170 GeV. Also the significance at 125 GeV  does not decrease (in-fact it marginally increases) when we go from $\xi=1$ to $\xi= 0.8$. This is again because of the above mentioned extra events from the decay of the scalar $H$ having $m_T< 125$ GeV that leak into the ``125 GeV signal".

 
The  Higgs boson at 125 GeV would also be seen in the $h\to ZZ\to 4l$ and $h \to \gÊ\g$ channels with  much better mass resolution. The cross-section can, however, not be measured accurately  with 15 fb$^{-1}$ data because of statistical uncertainties. This is because a 20 $\%$ reduction in the cross-section would be less than even a one sigma downward fluctuation. As far as the 170 GeV Higgs is concerned neither the $h\to ZZ\to 4l$ channel nor the  $h \to \gÊ\g$ channel  is sensitive to it with 15 fb data  for  $(1-\xi)=0.2$.

Thus we see that  in the scenario mentioned none of the measurements discussed so far would give any clear indication of the presence of the 170 GeV scalar.  The only difference between the $\xi=1$ and $\xi=0.8$ case would be in the shape of the significance vs $m_h$ curve, which is due to a difference in the underlying $m_T$-distribution. As can be seen from Fig.\ref{5fb} and Fig.\ref{10fb} the significance falls off much more sharply in the $\xi=1$ case. The $\xi=0.8$ curve lies within the two-sigma bands around the median $\xi=1$ expectation for low luminosities (see Fig.\ref{5fb}) and the difference in shape becomes significant only at higher luminosities  (see Fig.\ref{10fb}).

To disentangle the signal for $H$ one can treat signal due to a supposed SM Higgs at 125 GeV  as part of the background. The mass of the lighter Higgs can be  inferred from  excesses that would exist  in other channels like $h\to ZZ\to 4l$ and $h \to \gÊ\g$. This leads to a curve (Fig.\ref{sub} ) which peaks in the high mass region. For 10 fb$^{-1}$ luminosity we get almost a three-sigma excess  which indicates the presence of a heavier  Higgs boson  in addition to the Higgs at 125 GeV. Note that  we are subtracting the SM contribution for a 125 GeV Higgs whereas in reality the Higgs boson $h$ at 125 GeV has a reduced cross-section with $\xi=0.8$, and so the subtraction is unwittingly too large. The dashed curve in Fig.\ref{sub} shows the  significance curve if   the correct light Higgs contribution with  $\xi=0.8$ is subtracted from the signal and included in the background. Note that the peak position, even for the dashed curve is somewhat displaced to masses higher than 170 GeV. This is because of the falling background  at 170 GeV as discussed below eq.~Ê\ref{subtle}.

Finally let us comment on the CMS Higgs search analysis. Although both the ATLAS and CMS analyses have similar sensitivities, the CMS analysis is more involved as the cuts have been  optimized individually for each Higgs boson mass~\cite{cmsPAS}. The basic qualitative features that we have highlighted here, however,  should still be true for the CMS analysis. Even in the CMS study Higgs bosons  would show up as broad resonances in the $h/H \to WW\to 2l2\nu$ channel, in most of the  mass region considered here, before they are discovered in other channels with better mass resolution. Thus even in the CMS study the shape of the significance plots would be crucial for distinguishing an SM Higgs scenario from the case where there is an additional mixed-in scalar state. For the specific example we have considered even in the CMS study one expects that a heavier Higgs at 170 GeV, even with smaller couplings, would have substantial leakage of events to the signal window of a Higgs with lower mass and that the significance plot would have a longer tail if there is an additional heavier Higgs. To see how  our results still hold for the CMS study as far as the details are concerned, however, a thorough analysis needs to be done with the CMS cuts.


{\it Conclusions.} 
In this letter we have considered two mixed-in scalars having masses in the range where $h/H\to WW\to 2l2\nu$ channel is sensitive.  We find no dilution of significance of the `SM-like'   Higgs boson expected because of the mixing,   because of leakage of events from the other scalar into its signal region. Nevertheless,  with one extra mixed-in exotic Higgs boson, the shape of the  significance plot   for Higgs boson discovery in the $WW\to 2l2\nu$ channel -- even while performing search for one SM Higgs -- gets altered in a way that might reveal the existence of this other  Higgs boson.  The presence of the other scalar leads to a broadening of the excess over a larger mass range relative to the minmal SM Higgs case. In such a situation we propose that the second scalar can be more clearly identified by subtracting the contribution due to the `SM-like' Higgs

Of course, the total production rate for the `SM-like' Higgs, which could  be measured    in other channels like   $h \to \g\g$ and $h \to ZZ\to 4l$,  would be off compared to the SM in the event that the Higgs boson is mixed-in with a scalar.  The QCD uncertainties of production rate, and the statistical uncertainties that would be present in the initial phase of discovery would, however, be large enough that distortions in the $h \to WW\to 2l2\nu$ significance plot may be more revealing than simple accounting for the total rate.

\noindent
\emph{Acknowledgements:} We thank J. Zhu, A. Armbuster and A. Vartak for useful discussions.  This work is supported in part by the DOE Grant \#DE-FG02-95ER-40899 and by the European Commission under the contract ERC advanced grant 226371 �MassTeV�.

\begin{center}
\textbf{APPENDIX} 
\end{center}

In this appendix we review the procedure for evaluating exclusion confidence levels and discovery significances   assuming a Gaussian distribution for the expected number of  events. For exclusion of a particular  value of the mean expected signal $S$, the hypothesis being tested is the signal plus background hypothesis so that expected number of events, $N_{exp}$, has the mean value  $\bar{N}_{exp}=S+B$.  We assume a Poisson distribution for $N_{exp}$ with mean value $S+B$ and standard deviation $\sqrt{S+B}$. If the number of events finally observed in the experiment is $N_{obs}<S+B$ the signal plus background hypothesis is said to be excluded at 95$\%$ confidence level if the probability that $N_{exp}$ can fluctuate downward  form its mean value $S+B$ to a value less than or equal to $N_{obs}$ is less than 5$\%$.  For $S+B\gg 1$ the Poisson distribution we have assumed for $N_{exp}$ tends to a Gaussian distribution and the statement above  implies that signal values $S$, still allowed after setting the 95$\%$ CL~\footnote {Note that 95$\%$ CL corresponds to 1.96 standard deviations if both upward and downward fluctuations are considered and 1.64 standard deviations if only downward fluctuations are considered as is the case here.} bound would satisfy,
\bea
\frac{S+B-N_{obs}}{\sqrt{S+B}} &\leq& 1.64.
\eea
To find the median 95 $\%$ exclusion potential we take $N_{obs}=B$ to obtain,
\beq
 \frac{S}{\sqrt{S+B}}  \leq 1.64.
\label{exc}
\eeq
The upper limit on the  allowed signal is the maximum value of $S$ for which this condition holds.

The significance of a discovery, on the other hand, is defined as the significance for rejecting  the background-only hypothesis  if an excess  is seen over the background. We  assume a Poisson distribution with mean $B$ and standard deviation $\sqrt{B}$ for the background. The median discovery significance, $Z_0$ is then the number of standard deviations by which the background must fluctuate upward from its mean value to give an excess equal to the mean expected signal $S$, that is,
\beq
Z_0= \frac{S}{\sqrt{B}}.
\eeq
For a 5$\sigma$ discovery, for instance, we would have $Z_0=5$. The above expression, however, overestimates the significance if the statistics is low. A better approximation for the significance is given by the expression (defined as $S_{c12}$ in~\cite{Ball:2007zza}),
\beq
Z_0= 2(\sqrt{S+B}-\sqrt{B}).
\label{dis}
\eeq
This is the definition of significance we will use here.

Systematic uncertainties also play a very important role especially in the $h \to WW \to l \nu l \nu$ channel. The standard way to incorporate systematic effects is by convoluting the Poisson distribution for $N_{exp}$ (which is a Gaussian distribution in the large statistics limit) with the probability density function for the systematic uncertainty. Numerical convolution of the Poisson distribution with a systematic uncertainty having a Gaussian shape with standard deviation $\Delta B$ leads to the modification of eq.~\ref{exc} and eq.~\ref{dis} to~\cite{Ball:2007zza},
\bea
\frac{S}{\sqrt{S+B+ (\Delta B)^2}}  \leq 1.64  {\rm~and,}\\
Z_0 = 2(\sqrt{S+B}-\sqrt{B})\sqrt{\frac{B}{B+ (\Delta B)^2}}.
\eea
respectively.

\end{document}